\documentclass{article}
\usepackage{mrs2005,epsfig}
\begin{document} 
\title{DEEP SPECTROSCOPIC SURVEY OF LYMAN BREAK GALAXIES AT $Z\sim5$}

\author{Masataka Ando, Kouji Ohta}
\affil{Department of Astronomy, Kyoto University, Kyoto 606-8502, Japan}
\author{Ikuru Iwata}
\affil{Okayama Astrophysical Observatory, National
Astronomical Observatory, Kamogata, Okayama 719-0232, Japan}
\author{Masayuki Akiyama, Kentaro Aoki}
\affil{Subaru Telescope, National Astronomical Observatory of Japan, 650
North A'ohoku Place, Hilo, Hawaii 96720, USA}
\author{Naoyuki Tamura}
\affil{Department of Physics, University of Durham, Durham DH1 3LE, UK}

\begin{abstract} 
We report results of deep optical spectroscopy with Subaru/FOCAS
for Lyman Break Galaxy (LBG) candidates at $z\sim5$.
So far, we made spectroscopic observations for 24 LBG
candidates among $\sim 200$ bright ($z'<25.0$) LBG sample
and confirmed 9 objects to be LBGs at $z\sim5$.
Intriguingly, these bright LBGs show
no or a weak Ly$\alpha$ emission and 
relatively strong low ionization interstellar metal absorption lines.
We also identified 2 faint ($z'>25.0$) objects to be at $z\sim5$
with their strong (EW$_{\rm rest}>$20\AA) Ly$\alpha$ emission.
Combining our results with other spectroscopic observations of galaxies at
the similar redshift range, we found a clear luminosity
dependence of EW$_{\rm rest}$ of Ly$\alpha$ emission, i.e., the lack of
strong Ly$\alpha$ emission in bright LBGs.
If the absence of Ly$\alpha$ emission is due to dust absorption,
these results suggest that bright LBGs at $z\sim5$ are in dusty and 
more chemically evolved environment than faint ones. 
This interpretation implies that bright LBGs started star formation
 earlier than faint ones, suggesting biased star formation.
\end{abstract} 

\section{Introduction} 
A survey of galaxies at high redshift is a direct approach to understand
formation and evolution of galaxies.
In this decade, the color selection technique utilizing
blue UV continuum slope and Lyman break
of star-forming galaxies has found so-called Lyman Break
Galaxies (LBGs; e.g., \cite{Ste03})
at $z\sim3$ ($\sim$2 Gyr after big bang).
Then various studies have been revealing their detailed properties
such as UV luminosity function (UVLF) \cite{Ste99},
UV spectroscopic features (e.g., \cite{Shap03}), and 
stellar population (e.g., \cite{Saw, Pap01}) etc,
opening the new era to understand galaxies at high redshift.

As a next step, a survey for LBGs at higher redshift is required.
We focused on redshift 5, because it is 
$\sim$1Gyr (corresponding to the maximum age estimation of LBGs at $z\sim3$) 
earlier than $z\sim3$ and
the highest redshift at which the two-color (secure) Lyman Break
selection using standard optical bands can be applied.
We have obtained deep and wide $V$,
$I_C$ and $z^{\prime}$-bands images with Subaru \cite{Iye04} and
Suprime-Cam \cite{Miya} in/around the GOODS-N field and the J0053+1234 field
(\cite{Iwa03}; Iwata et al. 2005 in prep.)
and successfully constructed a largest, $\sim$1000 galaxies
with $z'<26.5$, sample of LBG candidates at $z\sim5$ among other
similar surveys (\cite{Lehn03, Ouchi04a, Dick04}).

Now we investigate properties of these LBGs at $z\sim5$.
Their statistical and photometric properties such as UVLF rest-frame UV
to optical color, etc. are presented and discussed in Iwata's
contribution in this proceedings (see also \cite{Iwa03}).
In this paper, we report the results of spectroscopic observations of
our LBGs at $z\sim5$ and discuss their UV spectroscopic features.
Throughout this paper, we adopt flat $\Lambda$ cosmology, $\Omega_M=0.3$,
$\Omega_{\Lambda}=0.7$, and $H_0=70$[km s$^{-1}$Mpc$^{-1}$]. 
The magnitude system is based on AB magnitude.

\section{Observations} 
We made optical spectroscopy for a part of our LBG sample in the GOODS-N
field and
the J0053+1234 field using multi-object-spectroscopy (MOS) mode of the Faint
Object Camera and Spectrograph (FOCAS; \cite{kashi}) attached to the
Subaru Telescope.
Spectroscopic targets were selected from photometric catalog of our
survey for LBGs at $z\sim5$.
Details of imaging observation and color selection are described in
\cite{Iwa03} and Iwata et al.(2005 in prep.).
Main spectroscopic targets are our LBG candidates brighter than 
$z'=25.0$ mag.
Since one mask of FOCAS MOS covers a 6$^{\prime}\phi$ aperture diameter
field of view, we designed MOS masks to contain main targets as many as
possible on each MOS field.
So far, we observed four MOS fields: three masks in
the GOODS-N field and one mask in the J0053 +1234 field which
contain 24 bright targets.
We also included faint LBG candidates ($z' \geq 25.0$ mag)
as many as possible in each mask.

Spectroscopic observations were made in 2003 and 2004 under a clear condition.
We used the grism of 300 lines/mm blazed at 7500\AA\ and
the SO58 order cut filter.
This setting gave wavelength coverage from 5800\AA\ to 10000\AA\
depending on a slit position on a mask.
The MOS slit widths were
fixed to be 0.$^{''}$8, giving a spectral resolution of R$\sim$700
which was measured by night sky emission.
An exposure time of each frame was 0.5 hours, and a total effective exposure
time was $5-6$ hours.
This exposure time was set to detect continuum feature of main targets.
Seeings during the observing runs were 
$\sim$0.$^{\prime\prime}6 - 0.^{\prime\prime}$8.

\section{Results} 
Among 24 main targets, we identified 9 objects to be LBGs at $z\sim5$.
Examples of resultant spectra are shown in Figure 1.
\begin{figure}  
\vspace*{1.25cm}  
\begin{center}
\epsfig{figure=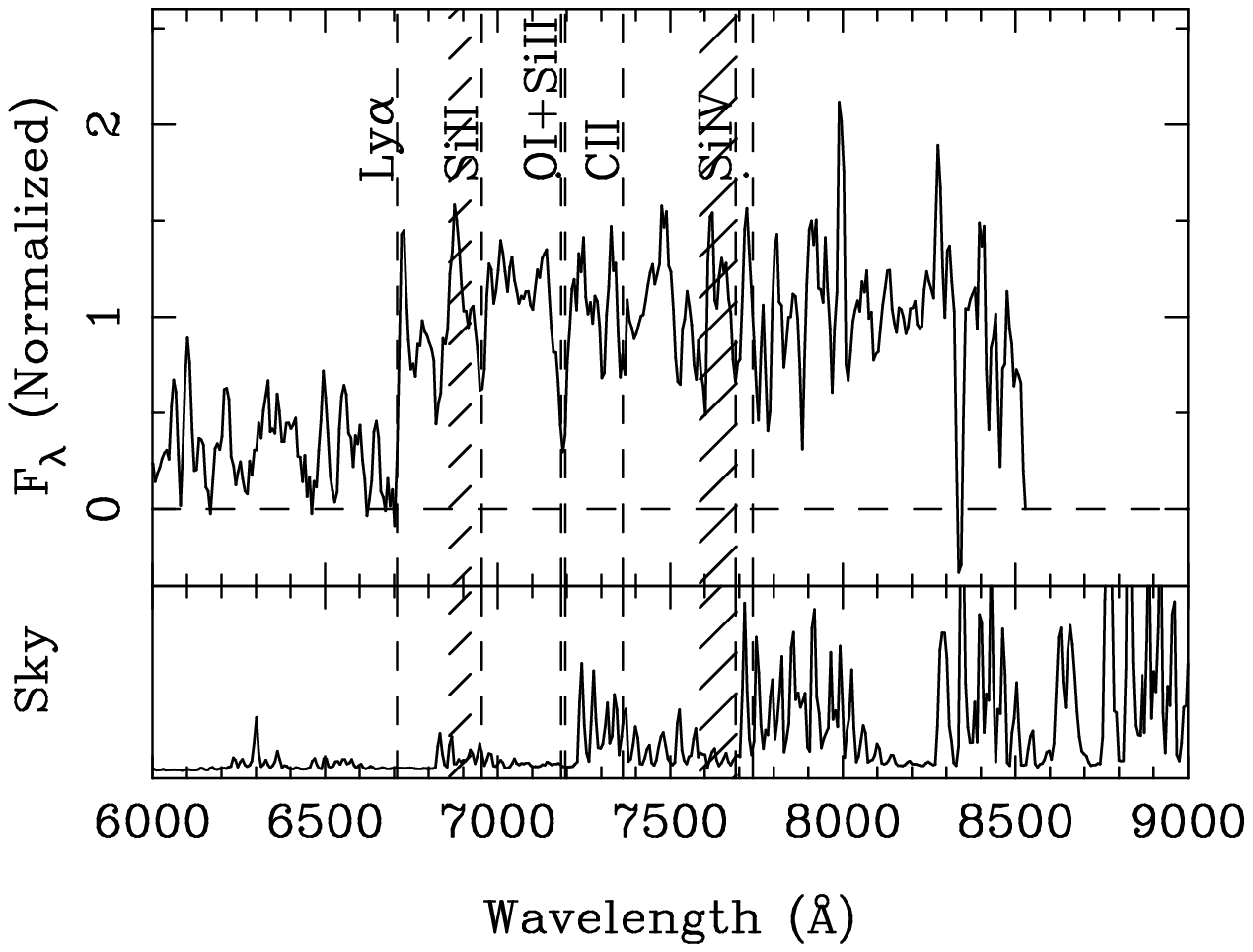,width=6.5cm}
\epsfig{figure=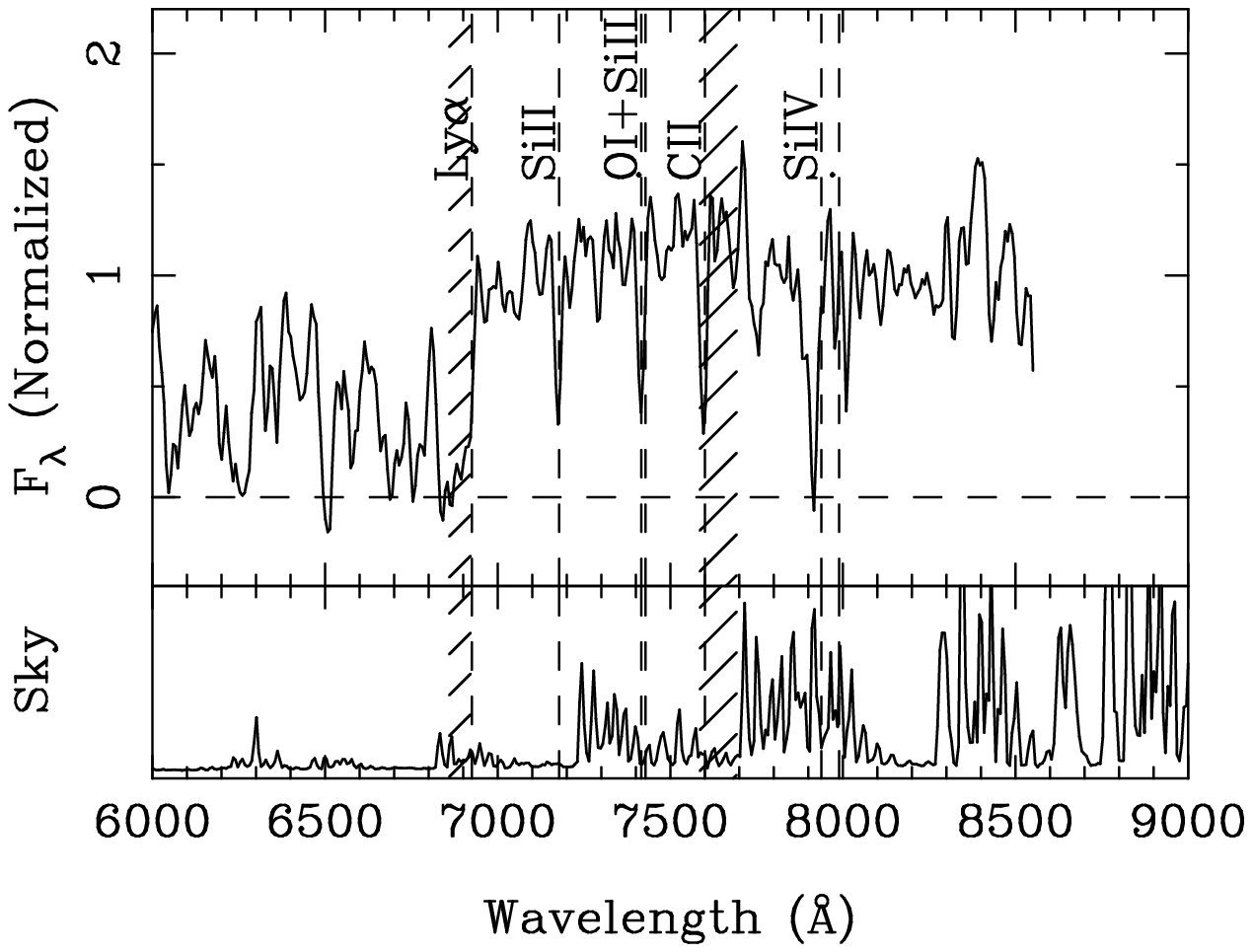,width=6.5cm}
\end{center}
\vspace*{0.25cm}  
\caption{
Examples of spectra of LBGs at $z\sim5$. Flux scale is $F_{\lambda}$ and
 normalized with the continuum level. Sky spectrum is shown in a lower
 panel of each figure, and atmospheric absorptions are shown as vertical
 hatched regions.
} 
\end{figure} 
Redshifts of these objects are confirmed with continuum depression
shortward of redshifted Ly$\alpha$ and some line features such as
Ly$\alpha$ emission and low
ionized interstellar (LIS) metal absorption lines which are
characteristic features of nearby starburst galaxies
(e.g., \cite{Hek98}) and LBGs (e.g., \cite{Shap03, Ste96a,frye}).
Intriguingly, these bright LBGs generally show
no or a weak Ly$\alpha$ emission and 
relatively strong LIS absorption lines, though the sample size is still small.
Figure 2 presents the composite spectra of bright LBGs at $z\sim5$
(thick line: \cite{Ando04}) and LBGs at $z\sim3$ (thin line:
\cite{Shap03}) for comparison.
\begin{figure}  
\vspace*{1.25cm}  
\begin{center}
\rotatebox{270}{\epsfig{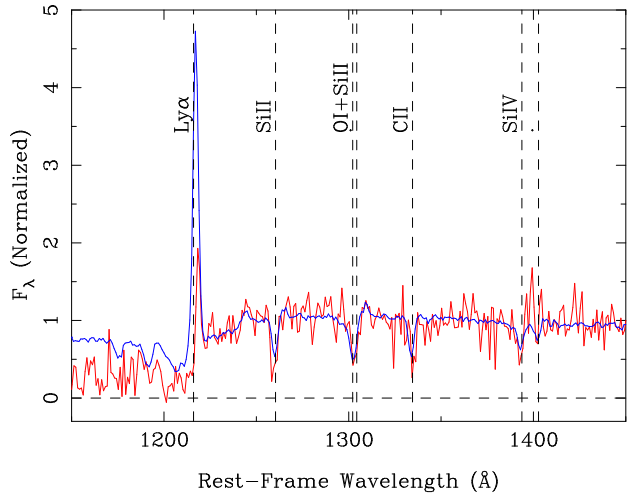}}
\end{center}
\vspace*{0.25cm}  
\caption{Composite spectrum of LBGs at $z\sim5$ (thick line;
 \cite{Ando04}) and that of LBGs at $z\sim3$ (thin line; \cite{Shap03}).
Main line features are shown as vertical dashed lines.
}
\end{figure} 
The average rest-frame equivalent widths of Ly$\alpha$ and three LIS
absorption lines (SiII $\lambda$1260,
OI+SiII $\lambda$1303, and CII $\lambda$1334) of
these bright LBGs at $z\sim5$ are $5.9$\AA\ and $-2.6$\AA, respectively.
The value of EW of Ly$\alpha$ emission is small
by considering that Ly$\alpha$ emission is often seen in LBGs at $z\sim3$
and more than 1/4 of them have strong 
(EW$_{\rm rest}>$20\AA) Ly$\alpha$ emission.
The average value of EW of three LIS absorption lines is stronger than
that of LBGs at $z\sim3$ ($-1.8$\AA: \cite{Shap03}).
Assuming the local relation between LIS absorption and metallicity by
\cite{Hek98}, we can estimate their metallicity of 12$+$log(O/H)$\sim8.0$
(1/5 solar).
We also found a velocity offset between the peak of Ly$\alpha$ emission
with respect to that of average of LIS lines; peaks of Ly$\alpha$
emission are redshifted $300 - 700$ km s$^{-1}$ to LIS absorption for
five objects.
Similar velocity offsets were also reported in LBGs at $z=3\sim4$
(e.g.,\cite{Shap03, frye}) which may be related to a large scale outflow.

Reminders of spectroscopic main sample are not identified because of
their low S/N, but we found 2 objects among them to be
possible elliptical galaxies at foreground redshift ($z\sim1$) from
their continuum features like 4000\AA\ break.

Besides main targets, we confirmed 2 objects among faint bonus targets with
$z'>25.0$ to be LBGs at $z\sim5$ using their strong and asymmetric
Ly$\alpha$ emission line, though LIS absorptions were not seen due to
low continuum S/N.
In contrast to the result for main, bright, targets describe above,
Ly$\alpha$ emissions of these
faint LBGs are quite strong (average EW$_{\rm rest}\sim$ 58.5\AA),
and these are expected to be detected as Ly$\alpha$ emitters (LAEs).
This result suggests that EW of Ly$\alpha$ emission of LBGs at $z\sim5$ 
depends on the UV luminosity.

Figure 3 shows the positions of identified objects in the
$V - I_C$ and $I_C - z'$ two color diagram.
Filled circles show LBGs at $z\sim5$ (7 bright objects as large
circles and 2 faint objects as small ones), and crosses show foreground
objects.
In order to examine our color selection criteria, we also observed 
some objects located outside but close to the criteria.
As a result, four objects were identified to be Galactic M stars which
are also plotted as crosses in Figure 3.
These results suggest that our selection criteria for LBGs at $z\sim5$
are reasonable.
\begin{figure}  
\vspace*{1.25cm}  
\begin{center}
\rotatebox{270}{\epsfig{figure=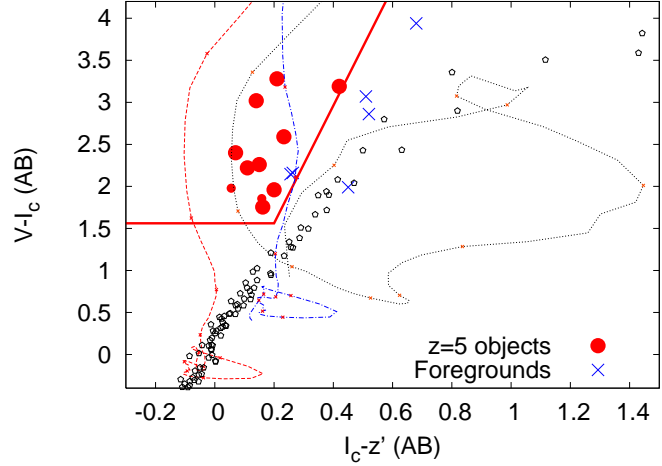,width=6.5cm}}
\end{center}
\vspace*{0.25cm}  
\caption{
Positions of identified objects in two-color diagram. Our color
 selection criteria \cite{Iwa03} for LBGs at $z\sim5$ 
are indicated by thick lines.
Filled circles represent the objects confirmed to be at $z\sim5$
(7 bright objects as big circles and 2 faint objects as small ones).
Crosses show objects identified to be foreground objects
 (Galactic M stars and possible ellipticals at $z\sim1$).
A dashed (a dot-dashed) line represents a color track of a model LBG
 spectrum with the $E(B-V)=0.0$ mag ($E(B-V)=0.4$ mag) from
 \cite{Iwa03}. A dotted line refers to a color track of an elliptical
 galaxy \cite{cww}. Small open pentagons indicate the colors of A0 -- M9
 stars calculated based on the library by \cite{Pick}.
} 
\end{figure} 

\section{Discussions} 
We found the sign of luminosity dependence of EW of Ly$\alpha$ emission in
LBGs at $z\sim5$; the lack of strong (EW$_{\rm rest}>$20\AA) Ly$\alpha$
emission in bright LBGs at $z\sim5$.
In order to examine this trend, we compiled past results of the
spectroscopy of galaxies at similar redshift.
Figure 4 shows the EW$_{\rm rest}$
of Ly$\alpha$ emission against the rest-frame UV absolute magnitude.
Filled circles show our results and 
filled squares are the results of spectroscopies of galaxies at $z=4.4-5.9$
\cite{Lehn03, Spi99, Wad99,Daw02, Dey98}.
We also show the SFR estimated from UV absolute
magnitude using the relation by \cite{Ken98}\footnote{In order to
derive UV absolute magnitudes and the SFR,
we assumed continuum slope $\beta$ of $-1$ which is a typical value of
LBGs at $z\sim3$.}.
This figure clearly shows that there are no UV luminous LBGs at $z\sim5$ with
strong (EW$_{\rm rest}>$20\AA) Ly$\alpha$ emission, while UV faint
ones tend to have strong Ly$\alpha$ emission.
In addition, there seems to be a UV magnitude threshold for LBGs with strong
Ly$\alpha$ emission around
$M_{1400}\sim-$21.5 mag which is almost the same as the $M_{\ast}$
magnitude of UV luminosity function of our $z\sim5$ LBG sample
\cite{Iwa03}.
In Figure 4, crosses show Lyman $\alpha$ emitters (LAEs) at $z\sim5.8$
from narrow-band imaging data \cite{Aji03}.
The EW distribution of LAEs is similar to that of faint LBGs with strong
Ly$\alpha$ emission, suggesting the fraction of LAEs to LBGs changes
with the UV luminosity at $z\sim5$ universe.
This is consistent with past result of \cite{Ouchi03}; a number ratio
of LAEs to LBGs at $z\sim5$ decreases with increasing UV luminosity.
\begin{figure}  
\vspace*{1.25cm}  
\begin{center}
\rotatebox{270}{\epsfig{figure=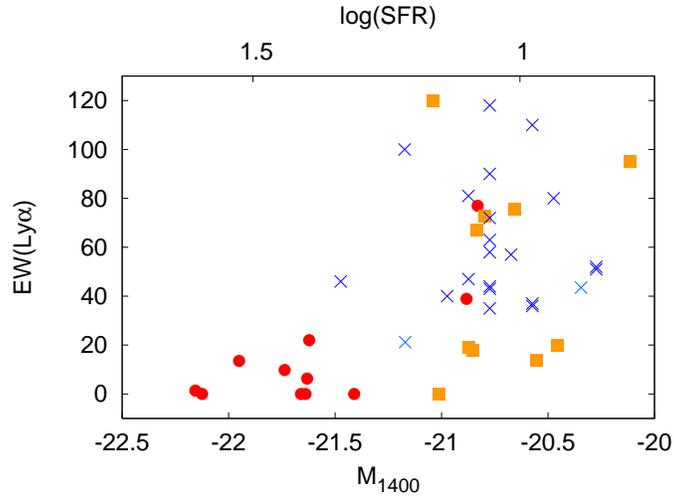,width=6.5cm}}
\end{center}
\vspace*{0.25cm}  
\caption{
Rest-frame EWs of Ly$\alpha$ emission vs. absolute
magnitude at rest-frame 1400\AA\ for galaxies at $z\sim5$.
Filled circles show our spectroscopic results and 
filled squares show results from \cite{Lehn03}
and serendipitously discovered objects at
$z\sim5$ \cite{Spi99, Wad99, Daw02, Dey98}.
Crosses represent LAEs at $z=5.8$ \cite{Aji03}
obtained from narrow-band imaging.
SFR estimated from UV absolute magnitude with the relation by
 \cite{Ken98} are also shown.
} 
\end{figure} 

If the absence of strong Ly$\alpha$ emission is due to the dust extinction,
luminous LBGs at $z\sim5$ may have chemically evolved to some extent.
Presence of strong LIS absorption and the estimated metallicity
($\sim$1/5 solar) also support this idea.
It seems that luminous LBGs at $z\sim5$ started star formation
relatively earlier than faint ones.
Further, results of clustering analysis of LBGs at $z\sim5$ show bright
LBGs have a larger correlation length than faint ones, suggesting
bright LBGs reside in more massive dark halos (\cite{Ouchi04b}; see also the
contribution by Iwata et al. in this proceedings).
This fact also implies that bright LBGs at $z\sim5$ have experienced
star formation earlier than faint ones, i.e., biased star formation in
the early universe.

Of course, velocity structure and distribution 
of HI gas (including dust) in/around the galaxy also affect the strength
of Ly$\alpha$ emission and its profile.
We found asymmetry of Ly$\alpha$ emission line and the velocity
offset between Ly$\alpha$ emission and LIS
absorption lines in a part of our sample,
which implies the presence of a large scale motion of
the neutral gas in/around LBGs at $z\sim5$.
Thus we can not rule out the possibility that the effect of gas geometry
and kinematics for the EW and profile of Ly$\alpha$ emission.

In any case, more spectroscopic sample of LBGs at $z\sim5$
is needed to study their spectroscopic features and discuss their
luminosity dependence presented in this paper,
which would give clues to understand evolution of galaxies in the early
universe.

%
 
\acknowledgements{ 
This work is based on data collected at Subaru Telescope which is
operated by the National Astronomical Observatory of Japan.
We are grateful to the FOCAS team, especially support astronomer
Youichi Ohyama, and all staffs of Subaru telescope
for their dedicated supports.
MAs are supported by a Research Fellowship of the Japan Society for the
Promotion of Science for Young Scientists.
}

\vfill 

\begin{thebibliography}{}{
\bibitem{Ste03}Steidel, C. C., Adelberger, K. L., Shapley, A. E., Pettini, M., Dickinson, M., \& Giavalisco, M. 2003, \apj, 592, 728
\bibitem{Ste99}Steidel, C. C., Adelberger, K. L., Giavalisco, M., Dickinson, M., \& Pettini, M. 1999, \apj, 519, 1
\bibitem{Shap03}Shapley, A. E., Steidel, C. C., Pettini, M., \& Adelberger, K. L. 2003, \apj, 588, 65
\bibitem{Saw}Sawicki, M., \& Yee 1998, \aj, 115, 1329
\bibitem{Pap01}Papovich, C., Dickinson, M., \& Ferguson, H. C. 2001, \apj, 559, 620
\bibitem{Iye04}Iye, M., et al. 2004, \pasj, 54, 833
\bibitem{Miya}Miyazaki, S., et al. 2002, \pasj, 54, 833
\bibitem{Iwa03}Iwata, I., Ohta, K., Tamura, N., Ando, M., Wada, S., Watanabe, C., Akiyama, M., \& Aoki, K. 2003, \pasj, 55, 415
\bibitem{Lehn03}Lehnert, M. D., \& Bremer, M. 2003, \apj, 593, 630
\bibitem{Ouchi04a}Ouchi, M., et al. 2004, \apj, 611, 660
\bibitem{Dick04}Dickinson, M., et al. 2004, \apj, 600, L99
\bibitem{kashi}Kashikawa, N., et al. 2002, \pasj, 54, 819
\bibitem{Hek98}Heckman, T. M., Robert, C., Leitherer, C.,Garnett, D. R., \& van der Rydt, F. 1998, \apj, 503, 646
\bibitem{Ste96a}Steidel, C. C., Giavalisco, M., Dickinson, M., \& Adelberger, K. L. 1996, \aj, 112, 352
\bibitem{frye}Frye, B., Broadhurst, T., \& Benitez, N. 2002, \apj, 568, 558
\bibitem{Ando04}Ando, M., Ohta, K., Iwata, I., Watanabe, C., Tamura, N., Akiyama, M., \& Aoki, K. 2004., \apj, 610, 635
\bibitem{Spi99}Spinrad, H., et al 1999, \aj, 116, 2617
\bibitem{Wad99}Waddington, I., Windhorst, R. A., Cohen, S. H., Partridge, R. B., Spinrad, H., \& Stern, D., 1999, \apj, 526, L77
\bibitem{Daw02}Dawson, S., et al., 2002, \apj, 570, 92
\bibitem{Dey98}Dey, A., Spinrad, H., Stern, D., Graham, J. R., \& Chaffee, F. H., 1998, \apj, 498, L93
\bibitem{Ken98}Kennicutt 1998, \araa, 36, 189
\bibitem{Aji03}Ajiki, M., et al. 2003, \aj, 126, 2091
\bibitem{Ouchi03}Ouchi, M., et al. 2003, \apj, 582, 600
\bibitem{Ouchi04b}Ouchi, M., et al. 2004, \apj, 611, 685
\bibitem{cww}Coleman, G. D., Wu, C.-C., \& Weedman, D. W. 1980, \apjs, 43, 393
\bibitem{Pick}Pickles, A.J. 1998, \pasp, 110, 863
} 
\end{thebibliography}
\end{document}